\documentclass{osa-article}

\journal{oe}
\begin{document}

\title{Integrated frequency comb laser with narrow intrinsic optical linewidth based on a dielectric waveguide feedback circuit}

\author{Jesse Mak\authormark{1,*}, Albert van Rees\authormark{1}, Youwen Fan\authormark{1}, Edwin J. Klein\authormark{2}, Dimitri Geskus\authormark{2}, Peter J. M. van der Slot\authormark{1}, and Klaus.-J. Boller\authormark{1}}

\address{\authormark{1}Laser Physics and Nonlinear Optics, MESA+ Institute for Nanotechnology, Department of Science and Technology, University of Twente, Enschede, The Netherlands\\
\authormark{2}LioniX International BV, Enschede, The Netherlands}

\email{\authormark{*}j.mak@utwente.nl} %% email address is required


\begin{abstract*}
%\LaTeX{} manuscripts submitted to OSA journals as of 30 May 2018 may use these instructions and this new single-column universal template format. Note that the final publishing format of OSA journals is not changing at this time, and authors will still need to adhere to article-length restrictions based on the final publishing format (which for some journals is two columns). Authors of Optics Letters articles and Optica letters and memoranda should continue using the legacy template for an accurate length check. Please note that OSA is no longer using OCIS codes.
We present an integrated hybrid semiconductor-dielectric (InP-Si\textsubscript{3}N\textsubscript{4}) waveguide laser that generates frequency combs at a wavelength around 1.5 $\mu$m with a record-low intrinsic optical linewidth of 34 kHz. This is achieved by extending the cavity photon lifetime using a low-loss dielectric waveguide circuit. In our experimental demonstration, the on-chip, effective optical path length of the laser cavity is extended to 6 cm. The resulting linewidth narrowing shows the high potential of on-chip, highly coherent frequency combs with direct electrical pumping, based on hybrid and heterogeneous integrated circuits making use of low-loss dielectric waveguides.
\end{abstract*}

%%%%%%%%%%%%%%%%%%%%%%%%%%  body  %%%%%%%%%%%%%%%%%%%%%%%%%%
\section{Introduction}
%Adherence to the specifications listed in this template is essential for efficient review and publication of submissions. Proper reference format is especially important (see Section \ref{sec:refs}).
Providing optical frequency comb sources in a chip-sized format is of high interest for a wide range of fields, for instance for dual-comb sensing\cite{CoddingtonOptica2016}, metrology\cite{UdemNature2002}, coherent optical communications\cite{PfeifleNatPhotonics2014}, and microwave photonics \cite{MarpaungLaserPhotonRev2013}. There are two central requirements regarding the optical coherence of the generated combs. The first is a highly equidistant spacing of the comb lines, i.e., a high mutual coherence of the comb lines, which shows up as narrowband beat frequencies of the comb lines. This requirement is usually fulfilled well via the nonlinear process that imposes mode-locking, e.g., four-wave mixing in Kerr combs \cite{SternNature2018} or saturable absorbers in passively mode-locked lasers \cite{WangLightSciAppl2017}, resulting in very narrow RF beat notes, e.g., in the sub-kHz range \cite{WangLightSciAppl2017}. The second requirement is that the spectral linewidth of the individual comb lines has to be extremely narrow, preferably in the kHz range or below. This corresponds to having a low frequency jitter of the comb. However, chip-based diode laser frequency combs, which are most attractive due to direct pumping with an electric current, usually fail to meet this requirement. This failure is due to a combination of intrinsic properties of semiconductor lasers, namely a short cavity length, high optical losses, and strong gain-index coupling \cite{Henry1982}. The resulting linewidths are typically tens of MHz in monolithically integrated diode lasers, e.g., indium phosphide (InP) lasers \cite{CheungIEEEPhotonTechnolLett2010}.

Providing a novel approach to overcome this linewidth limitation, i.e., to narrow the linewidth of the individual comb lines in chip-based diode lasers, is the main achievement of this paper. Microring resonators pumped by diode lasers \cite{SternNature2018} form an important alternative to generate frequency combs, based on Kerr comb generation. Although Kerr combs can ultimately provide a larger spectral coverage than diode laser combs, that approach will not be considered here due to its larger complexity, involving an additional optical pump threshold. Instead, we focus on the generation of a low-jitter comb with a diode laser that is directly pumped by an electric current.

A method which drastically improves the intrinsic linewidth of diode lasers (also named Schawlow-Townes limit \cite{SchawlowTownes1958}, fundamental linewidth or quantum linewidth), while maintaining a chip-sized format, is to extend the cavity photon lifetime \cite{FlemingMooradian1981} via hybrid or heterogeneous integration. In hybrid integrated lasers, a semiconductor amplifier is butt-coupled to a low-loss dielectric waveguide circuit, e.g., based on a silicon nitride (Si\textsubscript{3}N\textsubscript{4}) core with a silicon oxide (SiO\textsubscript{2}) cladding \cite{OldenbeuvingLaserPhysLett2013}, while in heterogeneous integrated lasers usually a III/V semiconductor is evanescently coupled to a low-loss waveguide circuit, e.g., based on silicon (Si) \cite{SantisPNAS2014}. Using these approaches, the linewidth of single-frequency semiconductor lasers was gradually reduced into the kHz range \cite{OldenbeuvingLaserPhysLett2013, SantisPNAS2014, DebregeasISLC2014, SantisCLEO2015}, and via hybrid integration with a Si\textsubscript{3}N\textsubscript{4} circuit even into the sub-kHz range (290 Hz) \cite{FanCLEO2017}. These findings suggest that extending this approach to multiple frequencies is a highly promising route to realize chip-based, narrow-linewidth frequency comb lasers. 

Chip-based frequency comb lasers with an extended cavity have already been demonstrated in the form of heterogeneous integrated mode-locked lasers \cite{SrinivasanFrontOptoelectron2014,SrinivasanIEEEJSelTopQuantumElectron2015,KeyvaniniaOptExpress2015,KeyvaniniaOptLett2015,WangLightSciAppl2017,DavenportPhotonRes2018}. The narrowest intrinsic linewidth, i.e., the lowest intrinsic comb jitter in a passively mode-locked integrated semiconductor laser reported so far, is 250 kHz with an InP-Si laser \cite{WangLightSciAppl2017}. However, that approach has a fundamental problem which can limit the achievable intrinsic linewidth. Namely, the cavity photon lifetime is extended using a waveguide circuit made of a low-bandgap material (silicon), having only a slightly larger bandgap (1.1 eV) than the generated light (0.8 eV). The low bandgap leads not only to a higher linear loss than in high-bandgap dielectrics \cite{HeckLaserPhotonRev2014}, but also to nonlinear losses due to two-photon absorption, which can deteriorate the achievable intrinsic linewidth\cite{VilenchikSPIE2015}. Compared to continuous-wave (CW) lasers, these nonlinear losses are expected to be large in mode-locked lasers, due to the relatively high intensity of the pulses that travel through the resonator, if it consists entirely of semiconductor material.

Here, we defeat such problems by extending the photon lifetime via low-loss, high-contrast dielectric waveguides. A most advanced platform with such properties is the Si\textsubscript{3}N\textsubscript{4}/SiO\textsubscript{2} waveguide platform, which provides bandgaps larger than 5 and 8 eV, respectively, low propagation losses down to 0.001 dB/cm, a high index contrast of 0.5, and exceptional maturity in terms of design and functionality \cite{RoeloffzenIEEEJSelTopQuantumElectron2018}.

In the following, we report a fully integrated, narrow intrinsic linewidth frequency comb laser based on dielectric waveguides. Using a hybrid integrated InP-Si\textsubscript{3}N\textsubscript{4} laser to generate frequency combs, we measure the linewidth of a single comb line by beating the output with a narrowband reference laser. We observe a low intrinsic linewidth of 34 kHz. This is, to our knowledge, much lower than reported for any fully integrated frequency comb laser so far. Although the comb currently comprises only a small number of lines (17) at a spacing of 5.3 GHz, this result confirms the strong potential for realizing on-chip, narrow-linewidth frequency combs via hybrid and heterogeneously integrated semiconductor lasers making use of low-loss dielectric waveguides.

\section{Methods}

\subsection{Laser description}
\begin{figure}[t!]
	\centering\includegraphics[width=12cm]{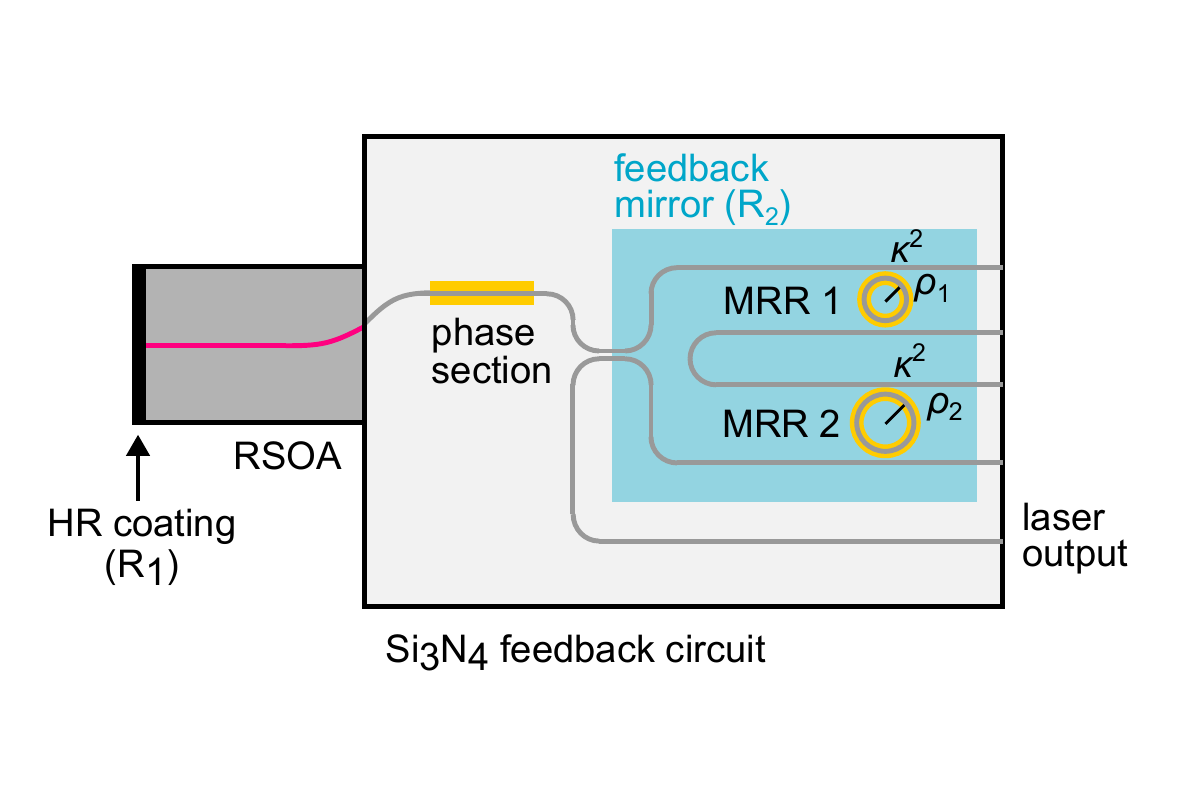}
	\caption{Schematic diagram of the hybrid waveguide laser (not to scale). The RSOA is equipped on its left-hand facet with an HR coating that has a power reflectivity $R_1$. The Si\textsubscript{3}N\textsubscript{4} circuit contains a phase section and a feedback mirror (blue tinted area) which has an effective power reflectivity $R_2$. This mirror comprises two sequential (Vernier) microring resonators (MRRs), denoted by MRR 1 and MRR 2, with radii $\rho_1$ and $\rho_2$. The power coupling between each of the rings and the corresponding straight waveguides is denoted by $\kappa^2$. The phase section and the MRRs can be tuned using resistive electric heaters, which are indicated in yellow.  }
	\label{FigCircuit}
\end{figure}
The hybrid laser investigated in this work is shown schematically in Fig. \ref{FigCircuit}. The laser cavity is formed using an InP quantum well based, double-pass reflective semiconductor optical amplifier (RSOA, fabricated by the Fraunhofer Heinrich Hertz Institute) and a Si\textsubscript{3}N\textsubscript{4} chip. The latter contains a circuit with two microring resonators (MRRs) that form a feedback mirror with adjustable spectral filtering. The Si\textsubscript{3}N\textsubscript{4} waveguides, having a symmetric double-stripe cross section\cite{WorhoffAdvOptTechnol2015}, offer a low propagation loss of about 0.1 dB/cm and a wide transparency range (400 nm to 2.35 $\mu$m).

Light is generated at a wavelength around 1.54 $\mu$m  in the RSOA, which is equipped on its back facet with a highly-reflective (HR) coating. At the InP-Si\textsubscript{3}N\textsubscript{4} interface, the waveguides are angled with respect to the facet normal, to minimize back reflections into the guided mode of the RSOA. In the Si\textsubscript{3}N\textsubscript{4} chip, the light first passes through a section where an extra optical phase can be imposed via the thermo-optic effect using a resistive electric heater, to fine-tune the spectral positions of the longitudinal laser cavity modes. Next, the light is guided through two sequential microring resonators (MRR 1 and MRR 2) with slightly different ring radii ($\rho_1$ = 136.5 $\mu$m and $\rho_2$ = 140.9 $\mu$m, respectively), which are thermo-optically tunable as well. The power coupling between each of the rings and the corresponding straight waveguides, $\kappa^2$, was designed to have a value of 10\%.

The purpose of the MRRs is twofold. First, the resonators increase the cavity photon lifetime, because they pass light in multiple roundtrips \cite{LiuApplPhysLett2001}, which lowers the intrinsic linewidth. Second, the MRRs form a spectral Vernier filter, which we use to induce the generation of a comb spectrum, as will be discussed in Section \ref{MethodsInducingCombGeneration}.

After passing through the MRRs, part of the light (20\% of the power) is returned to the InP amplifier using a directional coupler. The other part (80\% of the power) is guided to the edge of the Si\textsubscript{3}N\textsubscript{4} chip and forms the laser output.

The RSOA has a single-pass geometrical length $L_1$ of 700 $\mu$m, and a group index $n_1$ of about 3.6 at a wavelength of 1.5 $\mu$m. The effective \cite{LiuApplPhysLett2001} half-roundtrip length of the Si\textsubscript{3}N\textsubscript{4} feedback circuit (including multiple roundtrips in the MRRs), $L_{2}$, is 1.6 cm at the Vernier filter resonance. The group index of the feedback circuit, $n_2$, is 1.715. The corresponding effective roundtrip optical path length of the laser cavity, $L_{\mathrm{cav}}$, is about 6 cm.

To improve stability and reproducibility, the laser is hybrid integrated, i.e., the RSOA and Si\textsubscript{3}N\textsubscript{4} chips are aligned for optimum mode matching and then fixed permanently \cite{FanIEEEPhotonJ2016}. A single-mode polarization-maintaining fiber is bonded to the output waveguide of the Si\textsubscript{3}N\textsubscript{4} chip. The RSOA and the resistive heaters on the Si\textsubscript{3}N\textsubscript{4} chip are wire bonded and mounted inside a standard butterfly housing on a Peltier temperature controller. The laser is mounted on a printed circuit board with computer controlled voltage and current drivers.

\begin{figure}[t!]
	\centering\includegraphics[width=12cm]{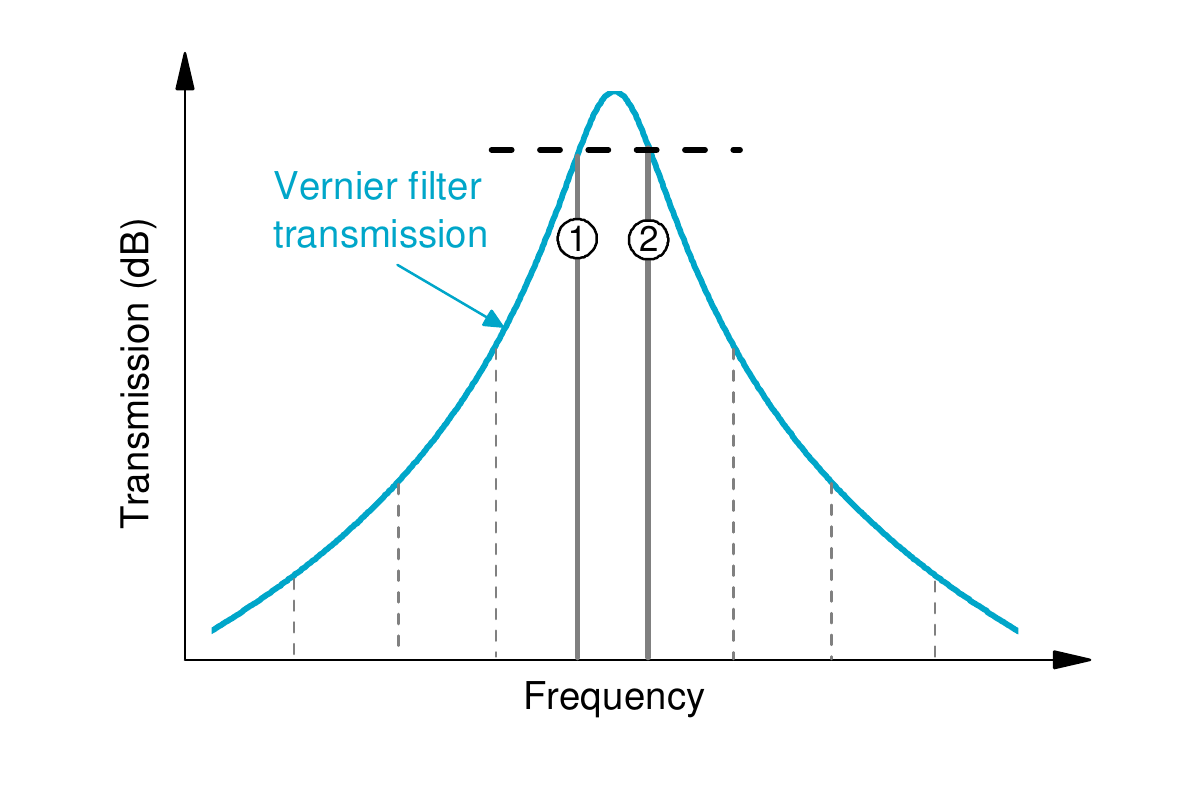}
	\caption{Schematic representation of the Vernier filter's transmission peak (solid blue line) and the spectral positions of the longitudinal laser cavity modes (gray lines). Positioning the laser modes as shown, with two central modes (labeled 1 and 2) receiving an equal amount of feedback (horizontal dashed line), is found to result in the generation of a frequency comb spectrum.}
	\label{FigOperationMethod}
\end{figure}

\subsection{Inducing comb generation}
\label{MethodsInducingCombGeneration}
Usually, Vernier filters and phase sections are tuned for the purpose of restricting laser oscillation to a single frequency by providing narrowband, strong feedback for only a single laser mode\cite{LiuIEEEPhotonTechnolLett2002, RabusIEEEPhotonTechnolLett2005, ChungIEEEPhotonTechnolLett2005}. However, multi-frequency comb generation, which is of interest here, can be imposed as well, by adjusting the spectral position of the two central, adjacent longitudinal modes of the laser cavity symmetrically around the filter’s transmission peak, via tuning the phase section. This spectral configuration is illustrated in Fig. \ref{FigOperationMethod}, which schematically shows the spectral shape of the filter's transmission peak as a solid blue line and the longitudinal modes of the laser cavity as gray lines. We found experimentally that frequency comb spectra are generated if the laser modes are tuned using the phase section, such that the two center modes receive an equal amount of feedback.

\subsection{Laser characterization}
\label{MethodsLaserCharacterization}
The laser output was characterized using an optical spectrum analyzer (OSA; Finisar WaveAnalyzer 1500S, 180 MHz resolution bandwidth). To characterize the output spectra with higher resolution, we recorded the mutual beating of the spectral lines, and the beating of single comb lines with an external single-frequency reference laser, using a fast photodetector (Discovery Semiconductors DSC30S, 20 GHz bandwidth) and an electrical spectrum analyzer (ESA; Agilent E4405B Spectrum Analyzer).

Of main interest is to measure the linewidth of a single spectral line, by recording beat notes between the hybrid laser and reference laser, because this forms a measure for the jitter of the diode laser comb. As the reference laser we used an extended cavity laser (SANTEC TSL-210), set to an output power of 3.2 mW. At this power, the reference laser has an intrinsic linewidth of 6 kHz, measured using the self-heterodyne technique \cite{OkoshiElectronLett1980}. To protect the hybrid laser and reference laser from undesired back reflections, the output of each laser was first sent through an optical isolator. Using a 2x2 fiber coupler, the light was superimposed and guided to the fast photodetector. To determine the intrinsic linewidth, we extracted the linewidth of the Lorentzian component in the RF beat spectrum, which determines the fall-off in the wings of the RF line. Near its center, the RF spectrum was found to be broadened with a near-Gaussian shape which is attributed to technical noise, such as from thermal or acoustic fluctuations and noise in the pump current, which are present in both the hybrid laser and the reference laser. This technical noise component can be removed with electronic frequency stabilization to reference cavities \cite{AlnisPhysRevA2008} or absolute frequency references \cite{NumataOptExpress2010}.
\begin{figure}[b!]
	\centering\includegraphics[width=12cm]{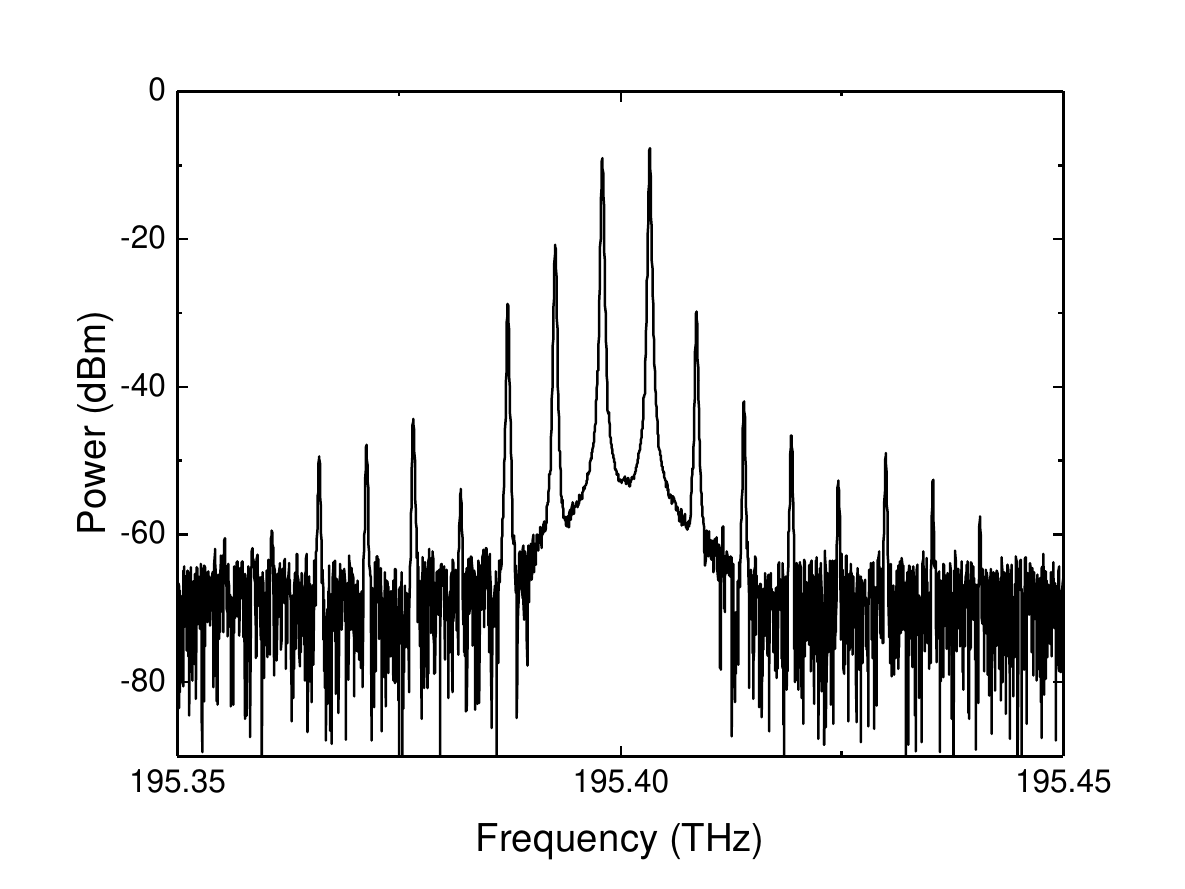}
	\caption{Measured frequency comb spectrum, using a pump current of 198 mA. The corresponding total output power is 2 mW.}
	\label{FigOpticalSpectrum}
\end{figure}
\section{Results and discussion}
\subsection{General laser behavior}
If the phase section is tuned such that one longitudinal laser mode coincides with the Vernier filter transmission peak, the laser shows single-frequency oscillation, with a high side-mode-suppression ratio of more than 50 dB. The output power is more than 10 mW at a pump current of 220 mA, and the laser threshold current is 12 mA. By temperature-tuning of the MRRs, the laser frequency can be tuned via mode hops over a wide spectral range of about 52 nm. However, when detuning the phase section such that two modes receive equal optical feedback, as depicted in Fig. \ref{FigOperationMethod}, the laser generates an optical comb spectrum. Fig. \ref{FigOpticalSpectrum} displays a representative example where 17 discrete, equidistant lines can be identified. The total output power is 2 mW. The power of the strongest line lies more than 50 dB above the noise level. The power of the spectral lines decreases strongly towards the edges of the spectrum, a feature that can likely be improved by using a filter with reduced steepness in the wings of its transmission spectrum. The frequency spacing of the lines is 5.3 GHz, which matches well with the expected free spectral range (FSR) of the laser cavity.

\begin{figure}[t!]	\centering\includegraphics[width=12cm]{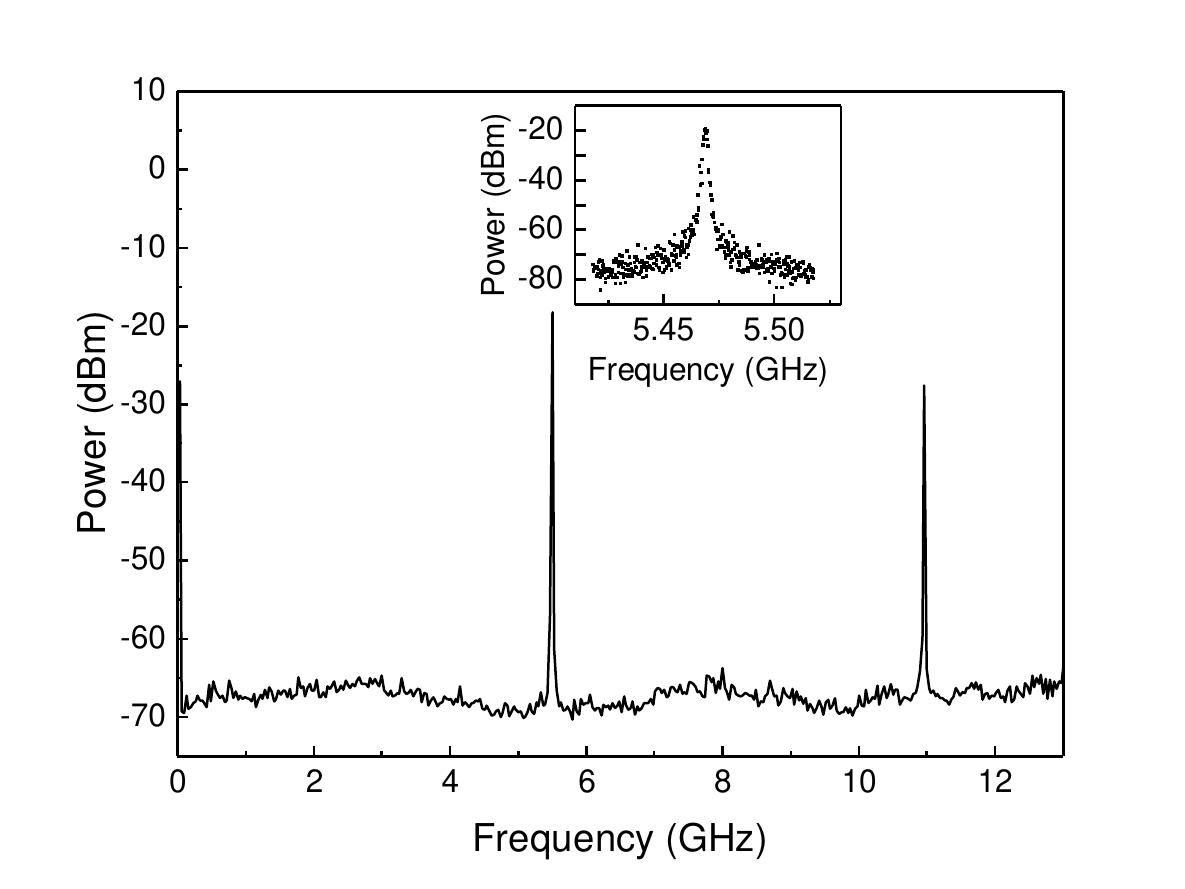}
	\caption{Measured RF beat spectrum using a pump current of 190 mA. The inset shows a close-up of the fundamental RF beat tone. A Voigt fit analysis of the spectrum shown at the inset yielded a Gaussian component with a 1.8 MHz linewidth and a Lorentzian component with an 18 kHz linewidth.}
	\label{FigRFLines}
\end{figure}

To inspect the mutual coherence of the comb lines, Fig. \ref{FigRFLines} shows the RF beat spectrum of the laser output for a pump current of 190 mA, recorded using a video bandwidth and resolution bandwidth of 3 MHz each, with a recording range of 13.2 GHz. Within the 13.2-GHz range of the RF spectrum analyzer, one can see two narrow peaks at about 5.5 GHz and 11 GHz. The two peaks lie more than 40 and 30 dB, respectively, above the noise level. A close-up of the fundamental peak, measured with a resolution bandwidth and video bandwidth of 1 MHz each, is shown in the inset. Analyzing this spectrum with a Voigt fit yields a very narrow Lorentzian component of 18 kHz, FWHM (technical noise Gaussian component is 1.8 MHz). The observation of a single RF peak at the FSR of the laser cavity and the low Lorentzian bandwidth of the RF peak indicate the presence of mode-locking. Such mode-locking without a saturable absorber has often been observed in quantum well lasers\cite{SatoIEEEJSelTopQuantumElectron2003, YangApplOpt2007, CaloOptExpress2015, RenaudierIEEEJQuantumElectron2007} due to spatial hole burning and four-wave-mixing via carrier density pulsations, which are driven by the beating of adjacent longitudinal modes \cite{DongPhysRevA2018}.

We note that due to the presence of resonators (the MRRs) in the laser cavity, the FSR is a function of frequency. We calculate that the FSR varies from 5 GHz at the resonance peak of the Vernier filter to 11 GHz (off-resonance). Nevertheless, both the optical and RF spectrum (Fig. \ref{FigOpticalSpectrum}-\ref{FigRFLines}) show that the measured spectral lines are equidistant. This provides direct confirmation that the laser is mode-locked, and is not simply a multi-mode laser oscillating at the center frequencies of the cavity resonances. Locking of modes to equidistant frequencies that can be relatively far off the cavity mode center frequencies is enabled by relatively high roundtrip loss (see Table \ref{TabParameters}). We have confirmed the described behavior via our numerical model \cite{FanOptExpress2017} based on \cite{VPI}, which predicts exactly equidistant output frequencies as well.

\subsection{Spectral linewidth measurements}
\begin{figure}[t!]
	\centering\includegraphics[width=12cm]{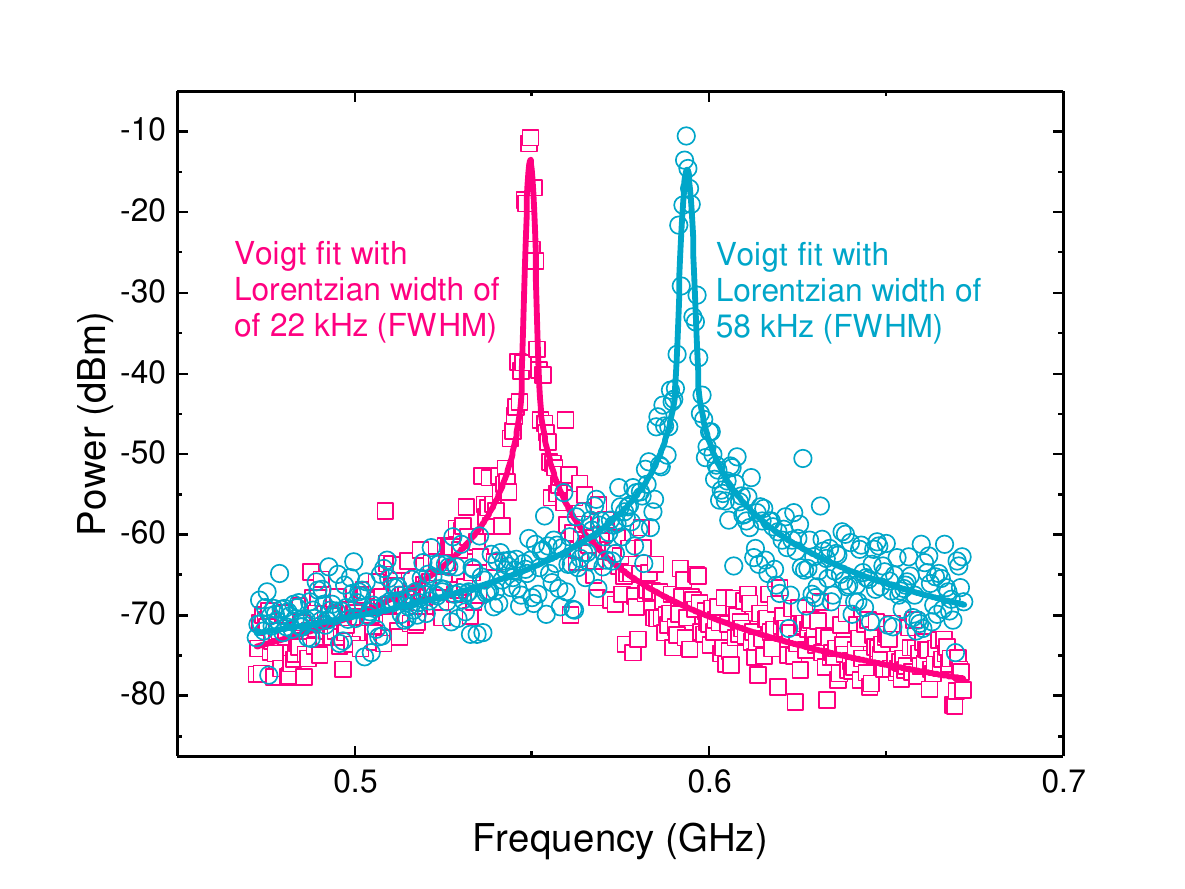}
	\caption{Measured beat note spectra (data points) obtained by beating the hybrid laser with a narrowband reference laser. Shown are two representative examples of the spectra, one with a broader (circles) and one with a narrower (squares) Lorentzian width. The Lorentzian components, obtained via a Voigt fit, amount to a 58 kHz and 22 kHz linewidth, respectively.}
	\label{FigLinewidth}
\end{figure}

To obtain quantitative information on the frequency jitter of the comb, we measured the intrinsic optical linewidth of the strongest comb line, by recording its RF beat note with a narrowband reference laser using the method described in Section \ref{MethodsLaserCharacterization}, with various settings of the hybrid laser’s pump current (up to 240 mA). In order to minimize RF line center broadening by technical noise, without having to implement an electronic frequency stabilization, we kept the time duration of the RF recordings short (4 ms). The narrowest beat notes were obtained using a pump current of 190 mA. A set of 40 beat note spectra was recorded sequentially, using a video bandwidth and resolution bandwidth of 1 MHz each. Each spectrum in this set was analyzed using a Voigt fit to extract the Lorentzian width of the spectra. Each spectrum showed a slightly different center position and spectral shape, due to drift of the hybrid laser and reference laser \cite{BandelOptExpress2016}. Two representative examples are shown in Fig. \ref{FigLinewidth}, which displays one broader beat note (Lorentzian width 58 kHz, FWHM) and one narrower beat note (22 kHz). The Lorentzian linewidth, obtained by averaging the Lorentzian components of the 40 spectra, is 40 kHz with a standard deviation of 20 kHz. To obtain the intrinsic linewidth of the hybrid laser we subtract the Lorentzian width of the reference laser (6 kHz), which yields a low intrinsic linewidth of approximately 34 kHz. To our knowledge, this linewidth is much lower than reported for any fully on-chip integrated frequency comb laser.
\begin{table}[]
	\caption{Parameters used for calculations}
	\label{TabParameters}
	\begin{tabular}{lll}
		\hline
		\textbf{Symbol} 	&
		\textbf{Value} 		& \textbf{Description} \\ \hline
		$n_{1}$				& 3.6         			& Group index of the RSOA                   						\\
		$n_{2}$				& 1.715              	& Group index of the external feedback circuit 						\\
		$\alpha_{1}$		& 1607 m$^{-1}$         & Internal loss coefficient of the RSOA            					\\
		$\alpha_2$			& 2.3 m$^{-1}$         	& Internal loss coefficient of the external feedback circuit    	\\
		$\beta$				& 0.9        			& Power coupling between the RSOA and the external feedback circuit	\\
		$n_{sp}$			& 2			   			& Spontaneous emission coefficient									\\   
		$\alpha$			& 3.0			   		& Linewidth enhancement factor										\\  
		$L_{1}$             & $700$ $\mu$m          & Single-pass geometrical length of the RSOA                		\\
		$L_{2}$				& 1.2 cm      			& Effective half-roundtrip length of the external feedback circuit 	\\
		$R_1$				& 0.9  			   		& Power reflection coefficient of the HR coating of the RSOA		\\
		$R_2$				& 0.2			   		& Power reflection coefficient of the external feedback circuit 	\\
		$\nu$				& 195.4 THz  			& Center frequency													\\
	\end{tabular}
\end{table}

To investigate whether the measured linewidth is consistent with theory, we estimated the expected linewidth of the hybrid laser. For mode-locked lasers without saturable absorber, as is the case here, the intrinsic linewidth of the individual modes is given by the Schawlow-Townes limit using the total average laser output power, as with single-frequency CW lasers \cite{HoIEEEJQuantumElectron1985, PaschottaApplPhysB2006}. Based on this argument, we use Henry's formula \cite{Henry1982} for the Schawlow-Townes linewidth of a semiconductor laser:
\begin{equation}\label{EqLinewidth}
\Delta\nu = \cfrac{v_g^2 h \nu n_{sp} g \alpha_m (1 + \alpha^2)}{8 \pi P_0}.
\end{equation}
Here, $v_g$ is the group velocity, $h$ is Planck's constant, $\nu$ the frequency of the light, $n_{sp}$ the spontaneous emission coefficient, $g$ the gain, $\alpha_m$ the mirror loss, $\alpha$ the linewidth enhancement factor, and $P_0$ the laser output power. As the laser consists of two sections (the RSOA and the dielectric feedback circuit), for the evaluation of Eq. (\ref{EqLinewidth}) we use an approach similar to \cite{KitaIEEEJSelTopQuantumElectron2014}, where $g$ is set equal to $\alpha_i + \alpha_m$, with $\alpha_i$ the internal loss, and using the following expressions for $v_g$, $\alpha_i$, and $\alpha_m$:
\begin{equation}
v_g = \cfrac{c}{ \left( \cfrac{L_{1} n_{1} + L_{2} n_{2} }{ L_{1}+ L_{2} } \right) },
\end{equation}
\begin{equation}
\alpha_i = \cfrac{L_{1} \alpha_{1} + L_{2} \alpha_{2} + \alpha_{c}}{L_{1} + L_{2}},
\end{equation}
\begin{equation}
\alpha_m = \cfrac{1}{L_{1} + L_{2} } \cdot \ln \left( \cfrac{1}{ \sqrt{R_1 R_2} } \right).
\end{equation}
Here, $\alpha_{c}$ is the coefficient accounting for the coupling loss at the InP-Si\textsubscript{3}N\textsubscript{4} interface, which is given by
\begin{equation}
\alpha_c = \ln \left( \cfrac{1}{ \beta } \right),
\end{equation}
with $\beta$ < 1 the power coupling efficiency between the InP and Si\textsubscript{3}N\textsubscript{4} waveguide. $R_1$ is the power reflection coefficient of the HR coating of the RSOA, and $R_2$ is the effective power reflection coefficient of the feedback mirror on the Si\textsubscript{3}N\textsubscript{4} chip.

A known limitation to the precision in evaluating Eq. (\ref{EqLinewidth}) is that several parameters, such as $n_{sp}$ and $\alpha$, need to be estimated. The parameters that were used here are listed in Table \ref{TabParameters}. These are the specified fabrication parameters, combined with parameters from our numerical study of a comparable hybrid laser \cite{FanOptExpress2017}. In choosing the value of $L_2$, we took into account that the central comb lines are slightly off-resonance with the Vernier filter, which decreases the effective half-roundtrip length of the external feedback circuit ($L_2 \approx$ 1.2 $\mathrm{cm}$). Inserting into Eq. (\ref{EqLinewidth}) an output power of 2 mW, as was measured during the experiments, and using the parameters of Table \ref{TabParameters}, we obtain a linewidth of about 16 kHz, which is of the same order as the measured value (34 kHz).

To briefly extrapolate whether the linewidth can be narrowed even further, we repeat the calculation with larger values of the extended resonator branch, $L_{2}$. When increasing this value to 15 cm, corresponding to an effective roundtrip optical path length $L_{\mathrm{cav}}$ of about 50 $\mathrm{cm}$ as in \cite{FanCLEO2017}, Eq. (\ref{EqLinewidth}) yields a linewidth of about 140 Hz. We thus expect that by extending the length of the external dielectric waveguide circuit, very low values for the intrinsic comb jitter, in the sub-kHz range, can be achieved.

\section{Conclusion}
We have presented a hybrid integrated InP-Si\textsubscript{3}N\textsubscript{4} laser that generates frequency combs with a record-low intrinsic linewidth of the comb lines of 34 kHz. This is achieved by extending the cavity photon lifetime using a low-loss Si\textsubscript{3}N\textsubscript{4} waveguide circuit, such that the effective, on-chip roundtrip optical path length is 6 cm. So far the comb spectrum contains lines at a spacing of around 5.5 GHz across a smaller bandwidth (17 lines). However, the spectral coverage may be improved and the line spacing may be reduced by modifying the dielectric feedback circuit. Examples are filters with reduced steepness in the wings of their transmission spectrum, and further extending the length of the laser cavity \cite{BautersOptExpress2011}. Our measured linewidth is more than a factor of 7 narrower than in previous work \cite{WangLightSciAppl2017}, and already approaches similar values as Kerr combs pumped by narrow-linewidth diode lasers\cite{SternNature2018}. The measured linewidth matches well with the expected theoretical value. Considering recent progress in reducing the linewidth of single-frequency semiconductor lasers \cite{FanCLEO2017}, it should be possible to reduce the linewidth of chip-based frequency combs much further, potentially into the sub-100-Hz range.

We note that the described approach is highly universal due to the wide transparency range of the dielectric waveguides. Therefore, comb generation should work equally well for other amplifier materials, including quantum dot amplifiers\cite{HuffakerApplPhysLett1998}, and may thus be used to construct lasers emitting at other wavelengths, including the visible range. An additional advantage is that, as the dielectric waveguide platform enables complex functionalities, the type of laser shown here is suitable for co-integration with other photonic integrated circuits, e.g., beam forming networks for microwave photonic applications\cite{RoeloffzenOptExpress2013}. These options show the important potential of electrically pumped frequency combs based on hybrid and hetereogeneous integration using low-loss dielectric waveguide circuits.

\section*{Funding}
This research is funded by the Netherlands Organization for Scientific Research (NWO) (Memphis II, 13537, Functional Hybrid Technologies).

\section*{Acknowledgments}
The authors would like to thank T. A. W. Wolterink, H. M. J. Bastiaens, C. G. H. Roeloffzen, and D. Marpaung for support and suggestions.

%%%%%%%%%%%%%%%%%%%%%%% References %%%%%%%%%%%%%%%%%%%%%%%%%

%%%%%%%%%% If using BibTeX:
%\bibliography{sample}

%%%%%%%%%% If preparing manually:

\end{document}